\documentstyle[aps,prl,twocolumn]{revtex}
\input epsf

\begin{document}
\draft
\title{Observation of the scissors mode and superfluidity of a trapped Bose-Einstein condensed gas}
\author{O.M.\thinspace Marag\`o, S.A.\thinspace Hopkins, J.\thinspace Arlt, E.\thinspace Hodby,
G.\thinspace Hechenblaikner, and C.J.\thinspace Foot.}
\address{Clarendon Laboratory, Department of Physics, University of Oxford,\\
Parks Road, Oxford, OX1 3PU, \\
United Kingdom.}
\date{\today}
\maketitle

\begin{abstract}
We report the observation of the scissors mode of a Bose-Einstein
condensed gas of $^{87}$Rb atoms in a magnetic trap, which gives
direct evidence of superfluidity in this system. The scissors mode
of oscillation is excited by a sudden rotation of the anisotropic
trapping potential. For a gas above $T_c$ (normal fluid) we detect
 the occurrence of oscillations at two frequencies, with the lower frequency
corresponding to the rigid body value of the moment of inertia.
Well below $T_c$ the condensate oscillates at a single frequency,
without damping, as expected for a superfluid.
\end{abstract}

\pacs{PACS numbers: 03.75.Fi, 05.30.Jp, 32.80.Pj, 67.90.+z}

The relationship between Bose-Einstein condensation (BEC)
\cite{Anderson} and superfluidity has been studied extensively in
liquid helium \cite{Tilley}, but only recently has it been
possible to examine it in condensates of dilute alkali metal
vapours \cite{Vortex,Ketterle}. Helium below its critical point is
described by a two-fluid model and the liquid is endowed with a
new degree of freedom, namely the relative motion between the
normal fluid and the superfluid. ``This degree of freedom is the
essence of the transport phenomena in He II known collectively as
superfluidity'' \cite{Huang}. In the dilute alkali metal
condensates various phenomena which imply the occurrence of
superfluidity have been observed e.g. collective modes of
excitation \cite{Jin} and demonstrations of the coherence of the
wavefunction \cite{coherence}. However in a recent theoretical
paper D. Gu\'ery-Odelin and S. Stringari \cite{GOS} describe how
the superfluidity of a trapped BEC may be demonstrated directly
and we report the results of such an experiment. Gu\'ery-Odelin
and Stringari analyse the so-called scissors mode in which the
atomic cloud oscillates with respect to the symmetry axis of the
confining potential and they point out that the scissors mode has
been used in nuclear physics to demonstrate the superfluidity of
neutron and proton clouds in deformed nuclei
\cite{Loiudice,Enders}.

\begin{figure}
\begin{center}\mbox{ \epsfxsize 3in\epsfbox{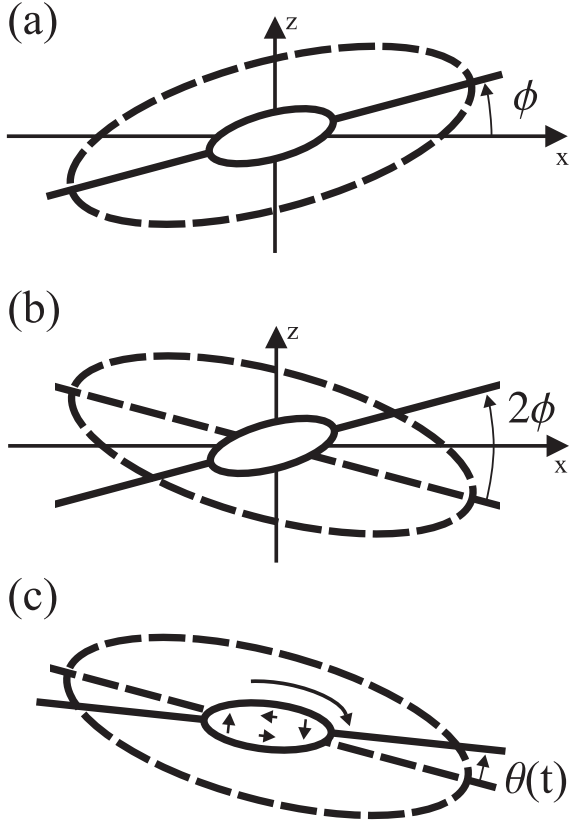}}\end{center}
\caption{The method of exciting the scissors mode by a sudden
rotation of the trapping potential. The solid lines indicate the
shape of the atomic cloud and its major axes. The dotted lines
indicate the shape of the potential and its major axes. (a) The
initial situation after adiabatically ramping on the field in z
direction, with cloud and potential aligned. (b) The configuration
immediately after rotating the potential, with the cloud displaced
from its equilibrium position. (c) The large arrow indicates the
direction of the scissors mode oscillation and the smaller arrows
show the expected quadrupolar flow pattern in the case of a BEC.
The cloud is in the middle of an oscillation period.(The angles
have been exaggerated for clarity.)}\label{ellipse}
\end{figure}

The full theoretical analysis of the scissors mode is given in
\cite{GOS} and we only outline the key points here. The starting
point is a BEC in an anisotropic harmonic potential with three
different frequencies $\omega_x \approx \omega_y < \omega_z$. The
scissors mode may be initiated by a sudden rotation of the
trapping potential through a small angle as indicated in
Fig.\ref{ellipse}. In the subsequent motion, the cloud is not
deformed provided that the change in the potential is too small to
excite shape oscillations. For a thermal gas both rotational and
irrotational fluid flow occur in the scissors mode and the normal
fluid is predicted to exhibit two frequencies corresponding to
those forms of motion. For the BEC there is only irrotational flow
because of its single valued wavefunction and therefore it only
exhibits one frequency, which is different from either of the
frequencies observed for the thermal cloud.

In our experiment the trapping potential is created by a
time-averaged orbiting potential (TOP) trap \cite{Petrich} which
is a combination of a static quadrupole field, of gradient
$B_{Q}^{^{\prime }}$ in the radial direction, and a time-varying
field

\begin{eqnarray*}
{\mathbf B}_T\left( t\right) =B_{r} \left( \cos \Omega t\ {\mathbf
e}_{x}+\sin \Omega t\ {\mathbf e}_{y}\right)+B_{z}\cos \Omega t\
{\mathbf e}_{z}.
\end{eqnarray*}

\noindent The term ${\mathbf B}_{z}\left( t\right) =B_{z}\cos
\Omega t\ {\mathbf e}_{z}$ is additional to the usual field of
amplitude $B_{r}$ rotating in the xy-plane. The effect of the
additional term  ${\mathbf B}_{z}\left( t\right)$ is to tilt the
plane of the locus of ${\mathbf B}=0$ by an angle $\xi
=\tan^{-1}(B_{z}/B_{r})$ with respect to the xy-plane. This causes
the symmetry axes of the potential to rotate through an angle
$\phi \approx \frac{2}{7} \xi $ in the xz-plane (this analytic
result is only valid for $\xi^2\ll1$). It is only necessary to
consider the xz-plane since the absorption image of the cloud is
projected onto this plane. Tilting the locus of ${\mathbf B}=0$
reduces the oscillation frequency in the z direction from its
value when $B_z=0$. Thus simply switching on ${\mathbf B}_z (t)$
also changes the cloud shape and so excites quadrupole mode
oscillations. To avoid this we first adiabatically modify the
usual TOP trap to a tilted trap and then quickly change ${\mathbf
B}_z (t)$ to $-{\mathbf B}_z (t)$. This procedure rotates the
symmetry axes of the trap potential by $2 \phi$ without affecting
the trap oscillation frequencies (Fig.\ref{ellipse}).

Our apparatus for producing BEC of $^{87}$Rb is described in
\cite{Arlt2} and only a few relevant features are outlined below.
We trap and cool the atoms in a small glass cell that forms part
of a differentially pumped system with two magneto-optical traps
(MOT)\cite{double}, one of which is a pyramidal configuration of
mirrors inside the vacuum system \cite{Arlt1}. The Helmholtz coils
which create the oscillating fields along the x, y and z
directions are driven by audio amplifiers and it was found to be
extremely important to filter out high frequency noise, especially
in the coils driving the field along the z-axis \cite{noise}. The
$^{87}$Rb atoms in the $F=2, m_F=2$ state, were probed by a 10
$\mu$s pulse of laser light propagating along the y-axis, resonant
with the $F=2, m_F=2\rightarrow F^{\prime}=3, m_F^{\prime}=3$
transition. The probe pulse was synchronized to the point in the
rotation when the magnetic field
points along the y-axis and this probing scheme is not affected by $\mathbf{B}%
_{z}\left( t\right) $. We are restricted to probing at times which
are multiples of the rotation period of the field (143 $\mu$s) but
this is very much smaller than the period of the oscillations in
the trap, as it must be for the TOP trap to work.

\begin{figure}
\begin{center}\mbox{ \epsfxsize 3in\epsfbox{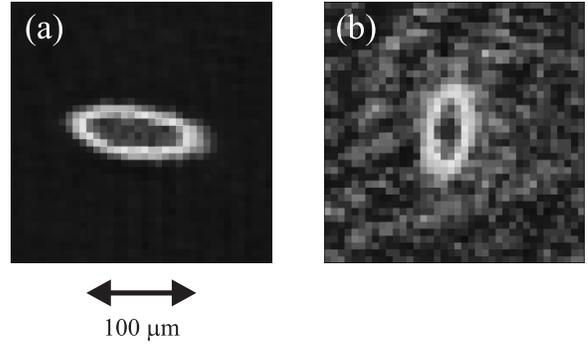}}\end{center}
\caption{(a) Image of the thermal cloud in the trap. (b) Image of
the condensate after 15 ms of time of flight. The condensate
expands most rapidly along the direction which is initially most
tightly confined, leading to a $90^{\circ}$ difference in
orientation as compared to the thermal cloud.}\label{pics}
\end {figure}

The following experimental procedure was used to excite the
scissors mode both in the thermal cloud and in the BEC. Laser
cooled atoms were loaded into the magnetic trap and after
evaporative cooling the trap frequencies were $\omega_x=90\pm 0.2$
Hz and $\omega_z=\sqrt{8}\; \omega_x$. The trap was then
adiabatically tilted in $1$ s by linearly ramping ${\mathbf
B}_{z}\left( t\right)$ to its final value, corresponding to $\phi
= 3.6^{\circ}$ and a reduction of the trap frequency $\omega_z$ by
$\sim 2 \%$. Suddenly reversing the sign of ${\mathbf B}_{z}\left(
t\right)$ in less than 100 $\mu$s excites the scissors mode in the
trapping potential with its symmetry axes now tilted by $-\phi$,
as described above. The initial orientation of the cloud with
respect to the new axis is $\theta_0=2 \phi$, so this angle is the
expected amplitude of the oscillations (Fig.\ref{ellipse}). The
angle of the cloud was extracted from a 2-dimensional Gaussian fit
of the absorption profiles such as those shown in Fig.\ref{pics}.

\begin{figure}
\begin{center}\mbox{ \epsfxsize 3in\epsfbox{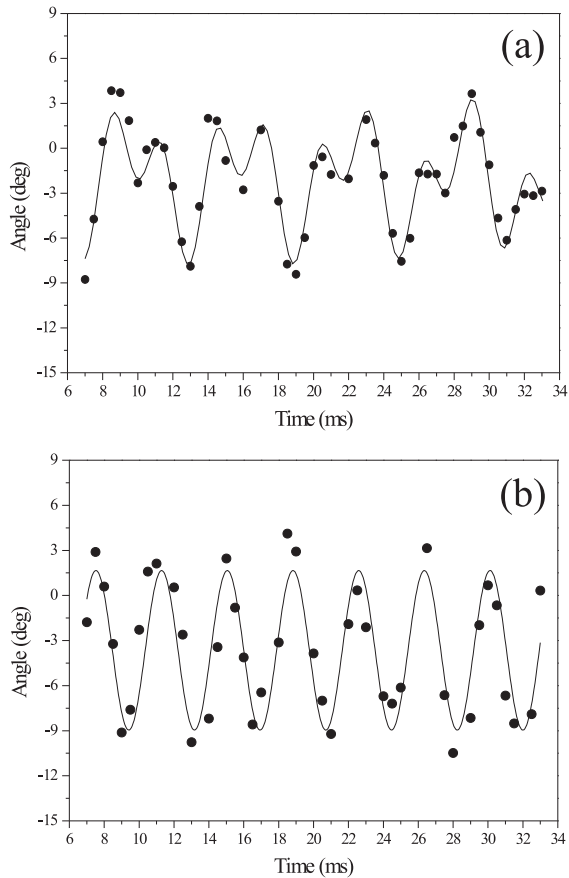}}\end{center}
\caption {(a) The evolution of the scissors mode oscillation with
time for a thermal cloud. For a classical gas the scissors mode is
characterized by two frequencies of oscillation. The temperature
and density of our thermal cloud are such that there are few
collisions, so no damping of the oscillations is visible. (b) The
evolution of the scissors mode oscillation for the condensate on
the same time scales as the data in (a). For the BEC there is an
undamped oscillation at a single frequency $\omega_c$. This
frequency is not the same as either of the thermal cloud
frequencies.}\label{scissors}
\end{figure}

For the observation of the thermal cloud the atoms were
evaporatively cooled to 1 $\mu$K which is about 5 times $T_c$, the
temperature at which quantum degeneracy is observed. At this stage
there were $\sim 10^5$ atoms remaining with a peak density of
$n_0\sim 2\cdot 10^{12}$ cm$^{-3}$. The scissors mode was then
excited and pictures of the atom cloud in the trap were taken
after a variable delay. The results of many runs are presented in
Fig.\ref{scissors}(a) showing the way the thermal cloud angle
changes with time. The model used to fit this evolution is the sum
of two cosines, oscillating at frequencies $\omega_1$ and
$\omega_2$. From the data we deduce $\omega_1/2\pi=338.5 \pm 0.8$
Hz and $\omega_2/2\pi=159.1 \pm0.8$ Hz. These values are in very
good agreement with the values $339\pm 3$ Hz and $159\pm 2$ Hz
predicted by theory \cite{GOS}; which correspond to
$\omega_1=\omega_z+\omega_x$ and $\omega_2=\omega_z-\omega_x$. We
measured $\omega_x$ and $\omega_z$ by observing the center of mass
oscillations of a thermal cloud in the untilted TOP trap and
calculating the modification of these frequencies caused by the
tilt. The amplitudes of the two cosines were found to be the same,
showing that the energy is shared equally between the two modes of
oscillation.

To observe the scissors mode in a Bose-Einstein condensed gas, we
carry out the full evaporative cooling ramp to well below the
critical temperature, where no thermal cloud component is
observable, leaving more than $10^4$ atoms in a pure condensate.
After exciting the scissors mode we allow the BEC to evolve in the
trap for a variable time and then use the time-of-flight technique
to image the condensate 15 ms after releasing it from the trap.
The repulsive mean-field interactions cause the cloud to expand
rapidly when the confining potential is switched off, so that its
spread is much greater than the initial size. The aspect ratio of
the expanded cloud is opposite to that of the original condensate
in the trap, so that the long axis of the time-of-flight
distribution is at $90^{\circ}$ to that of the thermal cloud as
shown in Fig.\ref{pics}. However this difference in the
orientation does not affect the amplitude of the angle of
oscillation. The scissors mode in the condensate is described by
an angle oscillating at a single frequency $\omega_c$

\begin{eqnarray}
\theta \left( t \right)=-\phi + \theta_0 \cos \left( \omega_c t
\right)\label{angle}
\end{eqnarray}

\noindent Figure \ref{scissors} (b) shows some of the data
obtained by exciting the scissors mode in the condensate.
Consistent data, showing no damping, was recorded for times up to
100 ms. From an optimized fit to all the data for the function in
Eq.\ref{angle} we find a frequency of $\omega_c/2\pi=265.6\pm 0.8$
Hz which agrees very well with the predicted frequency of $265\pm
2$ Hz from $\omega_c=\sqrt{\omega_x^2+\omega_z^2}$. The aspect
ratio of the time-of-flight distribution is constant throughout
the data run confirming that there are no shape oscillations and
that the initial velocity of a condensate (proportional to $\dot
\theta$) does not have a significant effect.

\begin{figure}
\begin{center}\mbox{ \epsfxsize 3in\epsfbox{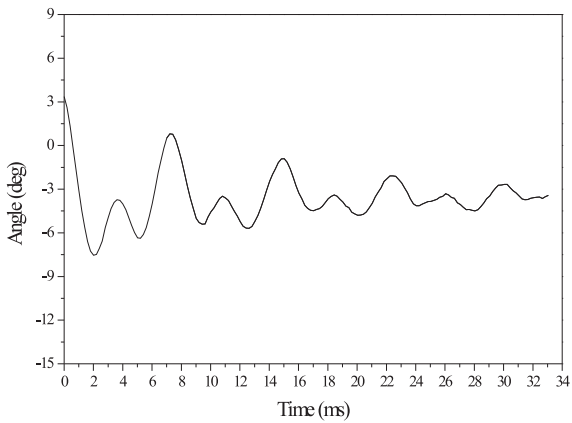}}\end{center}
\caption {A numerical simulation of
the scissors mode oscillation for a thermal cloud with a
temperature and density comparable to the condensate in our
experiment. The two frequency components are present and the
damping time is about 15 ms.}\label{damping}
\end{figure}

These observations of the scissors mode clearly demonstrate the
superfluidity of Bose-Einstein condensed rubidium atoms in the way
predicted by Gu\'ery-Odelin and Stringari \cite{GOS}. Direct
comparison of the thermal cloud and BEC under the same trapping
conditions shows a clear difference in behaviour between the
irrotational quantum fluid and a classical gas. Another
distinction is the lack of damping in the superfluid. The damping
of the classical gas is not apparent from our data because it is
not sufficiently dense and the time between collisions is many
oscillation periods. However the damping in a thermal cloud can be
calculated using the Direct Simulation Monte Carlo method
\cite{Wu}, for a mean density and temperature that are roughly the
same as those of the BEC. (Since the condensate and the thermal
cloud have different spatial distributions the comparison is only
approximate.) The results of such a numerical calculation for a
thermal cloud at a temperature of 90 nK and a peak density of $2
\cdot 10^{14}\ cm^{-3}$ are shown in Fig.\ref{damping} and the
plot shows that the scissors mode of a classical gas is strongly
damped under these conditions. Note that even for such high
densities both frequency components still occur in the thermal
cloud, giving behaviour that is clearly different from the
undamped, single frequency oscillation observed for the BEC. The
results of this numerical simulation show that the amplitude of
the lower frequency component is much smaller than that of the
higher frequency one. The higher frequency tends towards
$\omega_c$, the same frequency as the condensate, when the density
increases so that the hydrodynamic regime is reached (where there
are many collisions per oscillation period). However in this
regime the damping is so strong that only a few oscillation
periods would be observed as shown in \cite{GOS}.

In the near future we plan to measure the frequency of the
scissors mode at finite temperatures in the condensate i.e. where
a trapped thermal cloud is present in addition to the BEC. Under
these conditions the scissors mode should be damped \cite{GOS} in
a similar way to the quadrupole oscillations at finite temperature
\cite{Jin2}. There is still not good agreement between the
observed change in frequency of the m=0 oscillation mode of a
condensate at finite temperature and theoretical predictions
\cite{theory} and therefore using the scissors mode as an
alternative means of probing the interaction between the
condensate and thermal cloud is important. We have found that the
measurement of the angle of orientation of the cloud in the
scissors mode is quite robust and could be applied in situations
close to $T_c$ where only a small proportion of the atoms are
condensed. In this way the scissors mode should enable studies of
the relative motion of the thermal cloud and condensate
corresponding to motion of the superfluid through a normal fluid.

This work was supported by the EPSRC and the TMR program (No. ERB
FMRX-CT96-0002). O.M. Marag\`{o} acknowledges the support of a
Marie Curie Fellowship, TMR program (No. ERB FMBI-CT98-3077).

\begin{references}
\bibitem{noise} High frequency magnetic fields with components transverse to the
quantization axis drive transitions between Zeeman sub-levels and
were observed to shorten the magnetic trapping lifetime.
\end{references}
\end{document}